\documentclass[aps,twocolumn,superscriptaddress,showpacs,showkeys]{revtex4-1}
\usepackage{graphicx}
\usepackage[]{graphicx}
\usepackage{amsmath,amssymb,bm}
\usepackage{chemformula}

\begin{document}
\title{
Pareto front analysis
and 
multi-objective Bayesian optimization
for
($R$, $Z$)(Fe,Co,Ti)$_{12}$ ($R$ = Y, Nd, Sm; $Z$ = Zr, Dy)
}

\author{Taro \surname{Fukazawa}}
\email[E-mail: ]{taro.fukazawa@aist.go.jp}
\affiliation{CD-FMat, National Institute of Advanced Industrial Science
and Technology, Tsukuba, Ibaraki 305-8568, Japan}

\author{Takashi \surname{Miyake}}
\affiliation{CD-FMat, National Institute of Advanced Industrial Science
and Technology, Tsukuba, Ibaraki 305-8568, Japan}

\date{\today}
\begin{abstract}
We propose a scheme for investigating the correlation
and trade-off among 
target variables using a multi-objective Bayesian optimization (MBO).
We 
discuss the features of the Pareto front (PF)
of ThMn$_{12}$-type compounds,
($R$, $Z$)(Fe,Co,Ti)$_{12}$ ($R$ = Y, Nd, Sm; $Z$ = Zr, Dy)
in terms of 
magnetization, Curie temperature, and a price index
by using data from first-principles calculations,
and we extract the trade-off relations from the analysis. 
We show that 
the trade-off relationships can be used to determine
changes in the controllable variables
by using partial least squares regression.
For example, the tendency toward low cost and high Curie temperature is
related to the reduction in Dy and increase in Co.
We also discuss the efficiency of MBO
as a practical scheme to obtain the features of the PF.
We show that MBO can offer an approximated set for the PF
even when obtaining the true PF is difficult.
\end{abstract}
\pacs{TBD}
\preprint{Ver.0.5.6}
\maketitle

\section{Introduction}
Magnet compounds with the \ch{ThMn12} structure 
have attracted attention as a potential main phase
for high-performance permanent magnets
\cite{Ohashi88,Ohashi88b,Yang88,Verhoef88,
 DeMooij88,Buschow88,Jaswal90,Coehoorn90,Buschow91,Sakurada92,Sakuma92,Asano93,Akayama94,
 Kuzmin99,Gabay16,Koener16,Ke16,Fukazawa17,Fukazawa19}.
Hirayama et al. \cite{Hirayama15, Hirayama15b} synthesized a \ch{NdFe12} film and 
reported that the nitrogenated film exhibited a higher anisotropy field and magnetization
than \ch{Nd2Fe14B}, which is the main phase of the Nd magnet.
However, \ch{NdFe12} has not been found as a homogeneous bulk material.

In the development of functional materials, 
the substitution of elements in known materials
is a promising strategy.
Elements in \ch{$R$Fe12} ($R$: rare earth) could be substituted
to achieve higher performance and thermodynamic stability.
Hirayama et al. reported that introducing Co to a \ch{SmFe12} film increased the Curie temperature and magnetization at room temperature.
Ti stabilizes the \ch{ThMn12} structure, and bulk \ch{$R$Fe11Ti} can be synthesized \cite{Ohashi87,Ohashi88,DeMooij88}.
However, Ti doping greatly reduces the magnetization owing to substitution of Fe
and antiferromagnetic coupling between Ti and Fe.
Zr has also been studied extensively as a stabilizer that dopes the $R$ site,
which can avoid reduction of the magnetization
caused by substitution at the Fe site \cite{
Suzuki14,Sakuma16,Kuno16,Suzuki17}.

We also conducted several theoretical studies to search for
performance-enhancing
and stabilizing elements using first-principles calculations.
We investigated the formation energy of \ch{NdFe11$M$} from the unary phases
for $M$ = Ti--Zn and proposed that Co can
stabilize the \ch{ThMn12} structure \cite{Harashima16}.
We examined the stability of non-stoichiometric
doped systems of the \ch{ThMn12} phase,
where the phase competition with the \ch{Th2Zn17}-structure phase
was considered.
This result suggests that Co alone does not function as a stabilizer; 
however, co-doping with Zr and Co may stabilize
the \ch{ThMn12} structure \cite{fukazawa22b}.
For $R$-site doping, we have studied \ch{$R$Fe12} with
$R$ = La, Pr, Nd, Sm, Gd, Dy, Ho, Er, Tm, Lu, Y, and Sc
and proposed some rare-earth elements, including Dy, as possible
stabilizers \cite{Harashima18}.

Motivated by these works, in this study, we consider multiple doping, and we approach
the search for optimal dopants and their concentration as an optimization problem.
Machine learning techniques can deal well with 
this type of optimization problem, and several studies
have applied these techniques to materials exploration \cite{Ueno16,Ju17,Kikuchi18,Yuan18}.
For optimizing the chemical composition of non-stoichiometric systems,
we have been developing methods using Bayesian optimization 
combined with coherent potential approximation
in first-principles calculations \cite{Fukazawa19c,Fukazawa22}.
We have demonstrated that optimization with respect to 
chemical composition for a single target variable
can be performed efficiently with these techniques, and that
high-scored materials can be obtained with a 
small number of data acquisition processes.

In this paper, we consider a multi-objective optimization problem,
where multiple target variables are considered simultaneously.
In this type of problem, most of the data points are inferior to
other data points with respect to all target variables,
and these inferior data points are of much less interest.

The other important data points are called the Pareto front (PF).
We focus on the PF with respect to
the magnetization, the Curie temperature, and 
a price index as target variables 
in our analysis of first-principles data of
\ch{(R,Z)(Fe,Co,Ti)12} with the \ch{ThMn12} structure.
By using principal component analysis (PCA) and
partial least squares (PLS) regression, we determine how
the distribution of the PF is different from
that of all data points, and how
the information of the PF can be used to investigate
the trade-offs among the target variables.

However, 
in practical applications, obtaining the PF itself
can be difficult.
We discuss the use of multi-objective Bayesian optimization (MBO) for the problem,
and find that MBO is substantially more efficient than random sampling.
We also propose that MBO can be used as an efficient sampling method
for capturing the features of the PF.

\begin{figure}
    \centering
    \includegraphics[width=8cm]{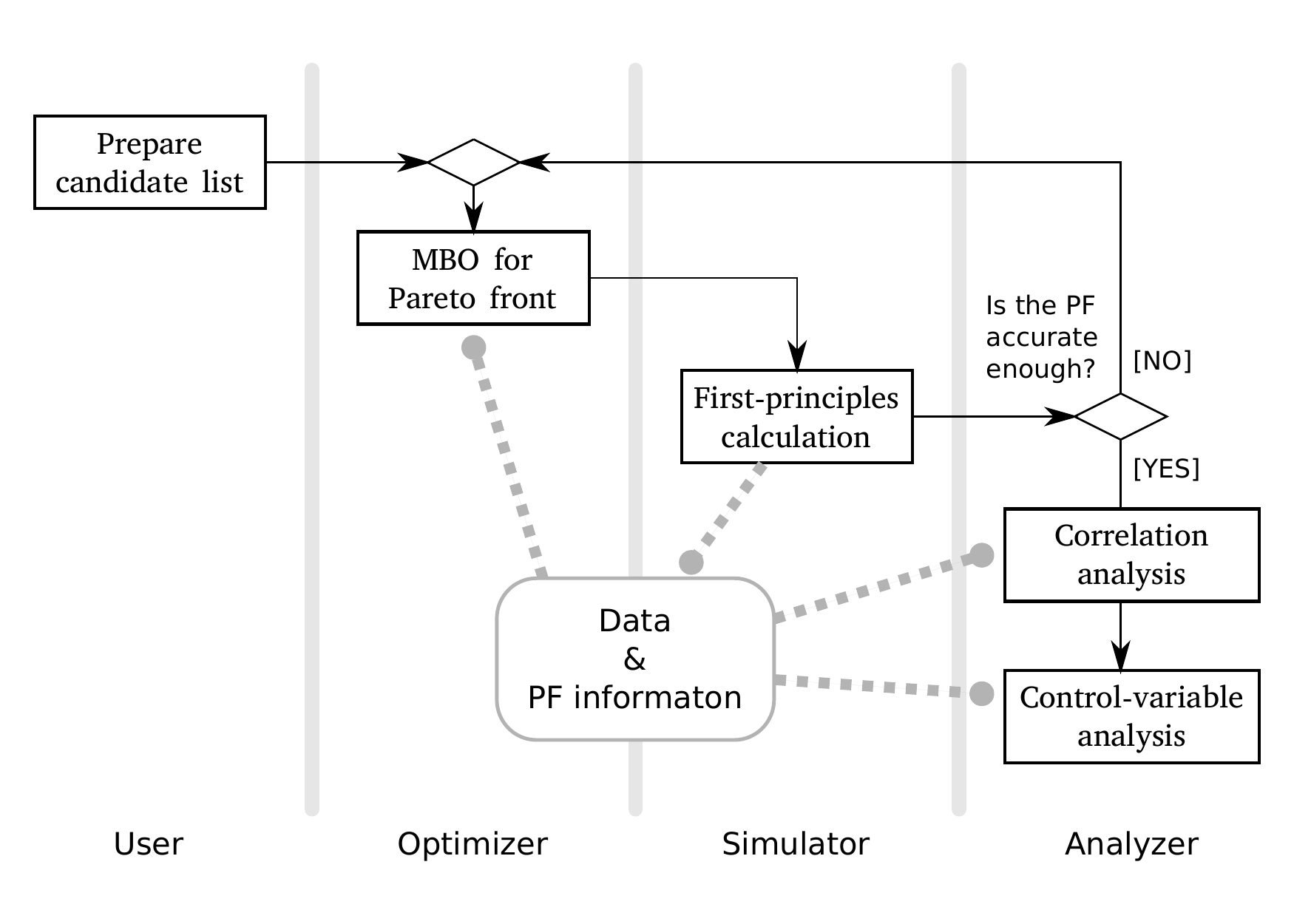}
    \caption{\label{flowchart}Work and data flow in the framework for 
    PF analysis.}
\end{figure}

Figure \ref{flowchart} summarizes our framework as a flow chart.
First, the user prepares a list of candidate materials
for which they want to obtain the PF.
Then, with the combination of a multiple-objective Bayesian optimizer
and a first-principles simulator, 
an approximate PF is efficiently obtained.
We discuss the accuracy and the efficiency of this part of the framework
in Section~\ref{SS:MBO}.
After an adequate PF is obtained,
correlation analysis is performed to visualize or quantify 
the trade-off between the target variables.
Control variable analysis can also be performed to determine 
how the target variables can be changed along the PF
by changing the controllable parameters.
As an example, in Section~\ref{SS:PFA}, 
we see the possibility that 
an acceptable compromise on the performance leads to 
a large reduction in the price,
and how the dopants and their 
concentration should be chosen.

\section{Methodology}
We use our data set for
\ch{(R_{$1-\alpha$}$Z$_{$\alpha$})(Fe_{$1-\beta$}Co_{$\beta$})_{$12-\gamma$}Ti_{$\gamma$}}
($R$ = Y, Nd, Sm; $Z$ = Zr, Dy; $\alpha$ = 0--1; $\beta$ = 0--1; $\gamma$ = 0--2)
obtained by first-principles calculations
based on density functional theory within the local density approximation \cite{Hohenberg64,Kohn65}.
The data set consists of 3630 data points.
We use AkaiKKR (MACHIKANEYAMA) in the data acquisition, which is based on the Korringa-Kohn-Rostoker
Green function method \cite{Korringa47,Kohn54}.
The lattice constants are determined by linear interpolation from those for \ch{$R$Fe12}, \ch{$R$Fe11Ti}, 
\ch{$Z$Fe12}, \ch{$Z$Fe11Ti}, and \ch{$R$Co12}.
The f-electrons in Nd, Sm, and Dy are treated as open cores \cite{Jensen91,Locht16,Richter98},
and the self-interaction correction \cite{Perdew81} is applied.
In treating the randomness from the doping, we use coherent potential approximation \cite{Soven70,Shiba71}.

Magnetization values are calculated from the magnetic moment and the volume. The Curie temperatures
are obtained within the mean field approximation from the Heisenberg Hamiltonian, the coupling parameters of which are determined by Liechtenstein's formula \cite{Liechtenstein87}.
We fix the price indices for the elements to 0.24 for Y, 6.3 for Nd, 0.28 for Sm, 0.0056 for Zr,
35 for Dy, 0.0056 for Fe, 4.8 for Co, and 0.61 for Ti,
which roughly represent their costs in arbitrary units.
The price indices for materials are calculated by the linear combination of the atomic price indices 
with the coefficient of the compositional ratio.

We use scikit-learn \cite{scikit-learn} for data analysis with PCA \cite{Wold87}.
and PLS regression \cite{Holcomb97}.
For Bayesian optimization, we use PHYSBO \cite{PHYSBO,Motoyama21} with the dimension of the random feature maps
set to 20, hypervolume-based probability of improvement \cite{Couckuyt14} used as an acquisition function,
and the number of the initial random sampling set to 10.

\section{Results and discussion}
\subsection{PF analysis}
\label{SS:PFA}
\begin{figure}
    \centering
    \includegraphics[width=8cm]{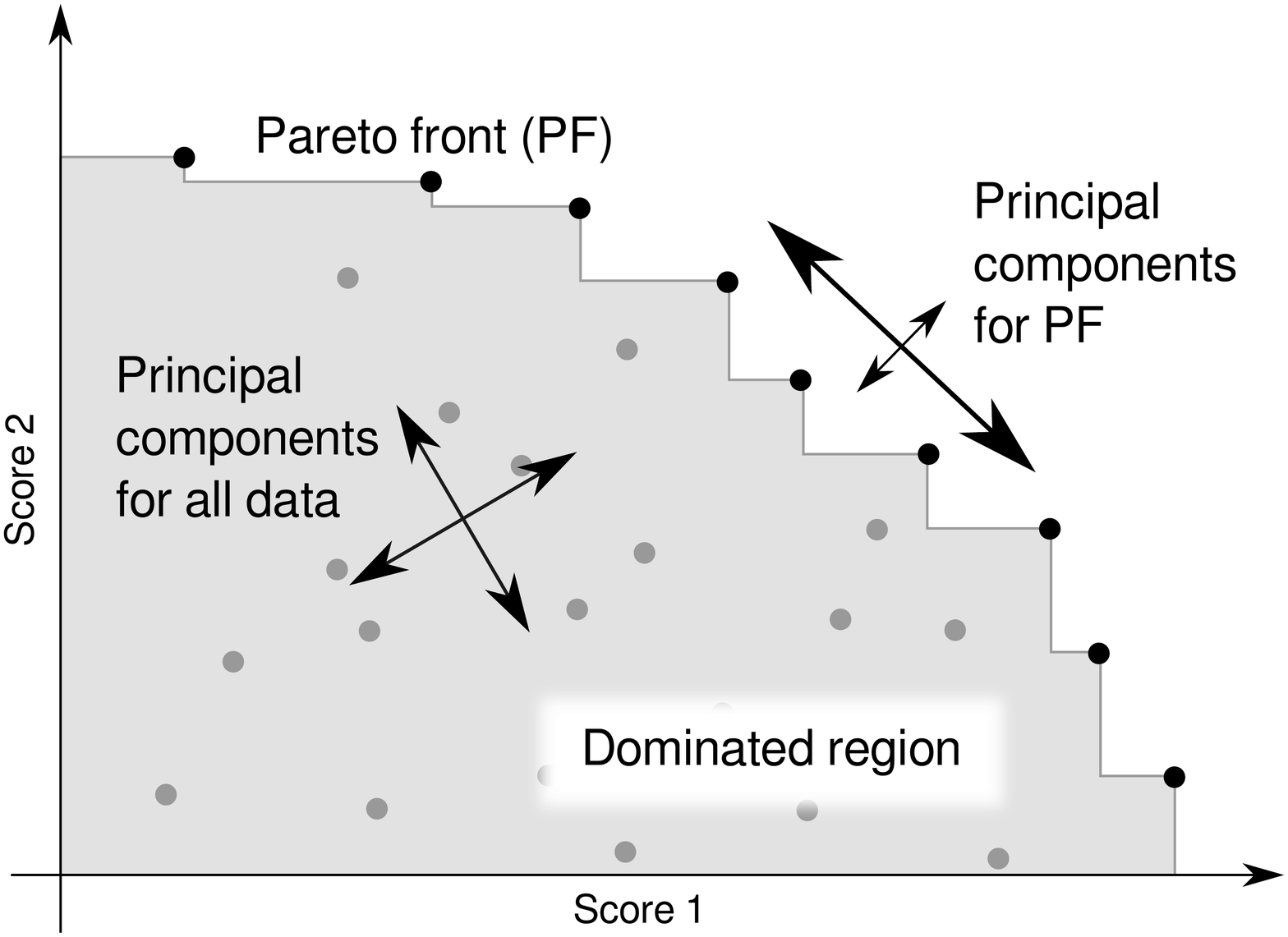}
    \caption{\label{Fig.PF}Schematic of the PCA
    with the PF.
    The arrows in the figure describe the PCs of the data
    and the variance of the data along the axes.}
\end{figure}

We extract the PF
with respect to the magnetization ($\mu_0M$),
Curie temperature ($T_\mathrm{C}$), and price index ($p$)
from the 3630 data points
by considering the following domination order.
When these properties for material A
are simultaneously superior to those for material B, that is, 
\begin{align}
    \mu_0M(\mathrm{A}) &\geq \mu_0M(\mathrm{B}), \label{pareto_1}\\
    T_\mathrm{C}(\mathrm{A}) &\geq T_\mathrm{C}(B), \label{pareto_2}\\ 
    p(\mathrm{A}) &\leq p(B), \label{pareto_3}
\end{align}
we say "A dominates B", and denote it as $\mathrm{A} \succeq \mathrm{B}$.
By introducing order $\succeq$,
the data set becomes a partially ordered set,
and the set that consists of the maximal elements is called the PF.
In other words,
if a data point has an advantage over all other points 
with at least one target variable, the data point is a member of the PF.
A schematic of an example is given in Fig.~\ref{Fig.PF},
where there are two target variables, score 1 and 2.
The nine closed circles denote the members of the PF in a case of two-dimensional
target variables. The gray area is the region dominated
by the PF.

\begin{figure}
    \centering
    \includegraphics[width=8cm]{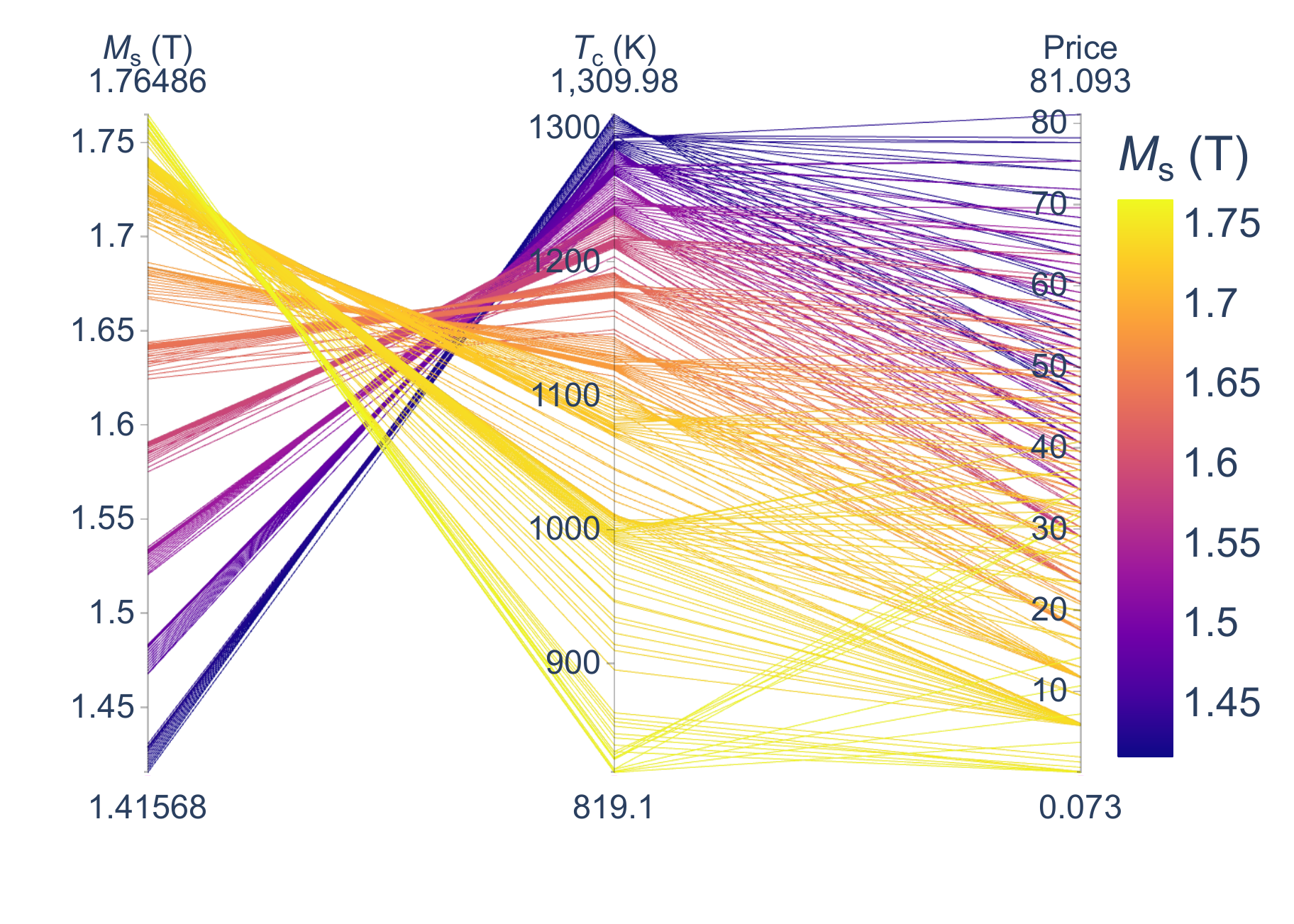}
    \caption{Parallel coordinate plot for the PF.\label{Fig.Parallel}
    The $M_{\mathrm s}$ axis shows the magnetization,
    the $T_{\mathrm C}$ axis shows the Curie temperature,
    and the Price axis shows the price index.}
\end{figure}

Correlation analysis follows
collection of information about the PF
(the optimizer-simulator loop) (Fig. \ref{flowchart}).
We use the results from the exhaustive simulation 
and show the data as
a parallel coordinate plot
for the target variables in the PF
in Fig.~\ref{Fig.Parallel}
to visualize the correlation among the target variables.
In the figure, a line denotes a system and 
links its target values on the parallel coordinates.
There is a strong correlation between the magnetization and 
Curie temperature, which is shown as sharp diagonal line bunches 
between the $M_{\mathrm s}$ (magnetization) and
$T_{\mathrm C}$ (Curie temperature) axes.
In contrast,
the price index has a weaker 
correlation with the magnetization and Curie temperature.
Thus,
the price can be reduced greatly
by accepting a small reduction in the 
magnetization and Curie temperature.

\begin{figure}
    \centering
    \includegraphics[width=8cm]{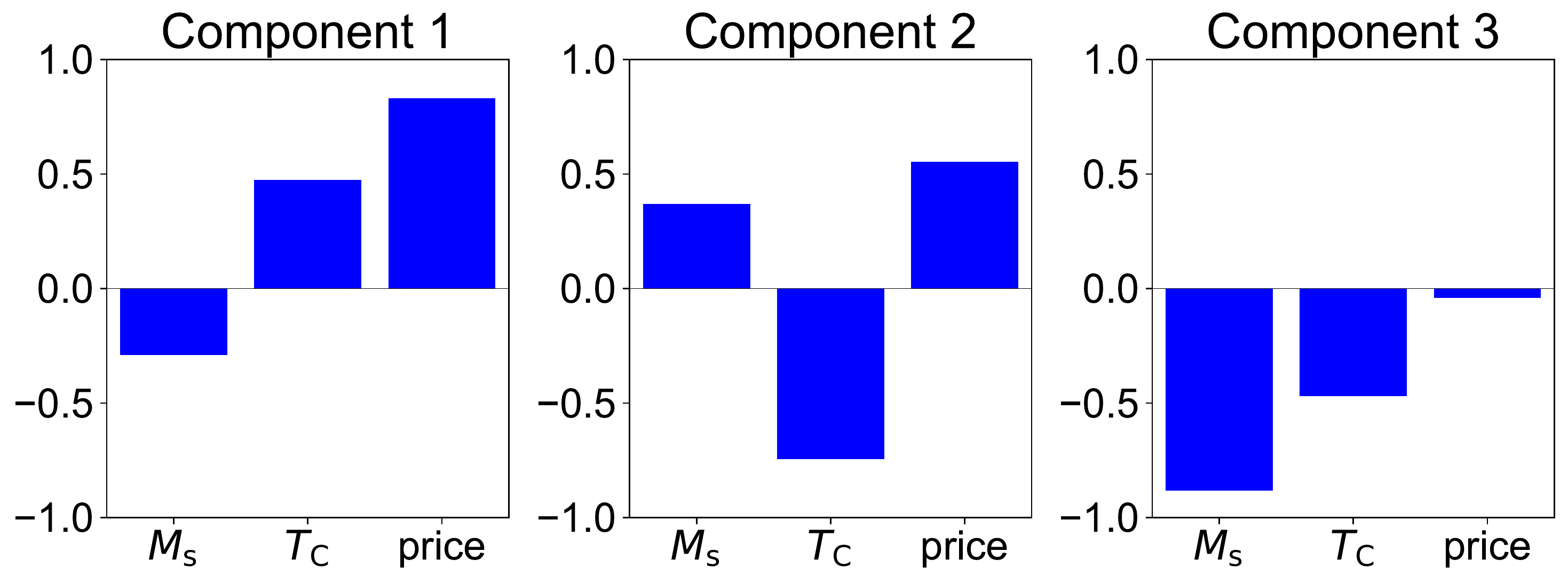}
    \includegraphics[width=8cm]{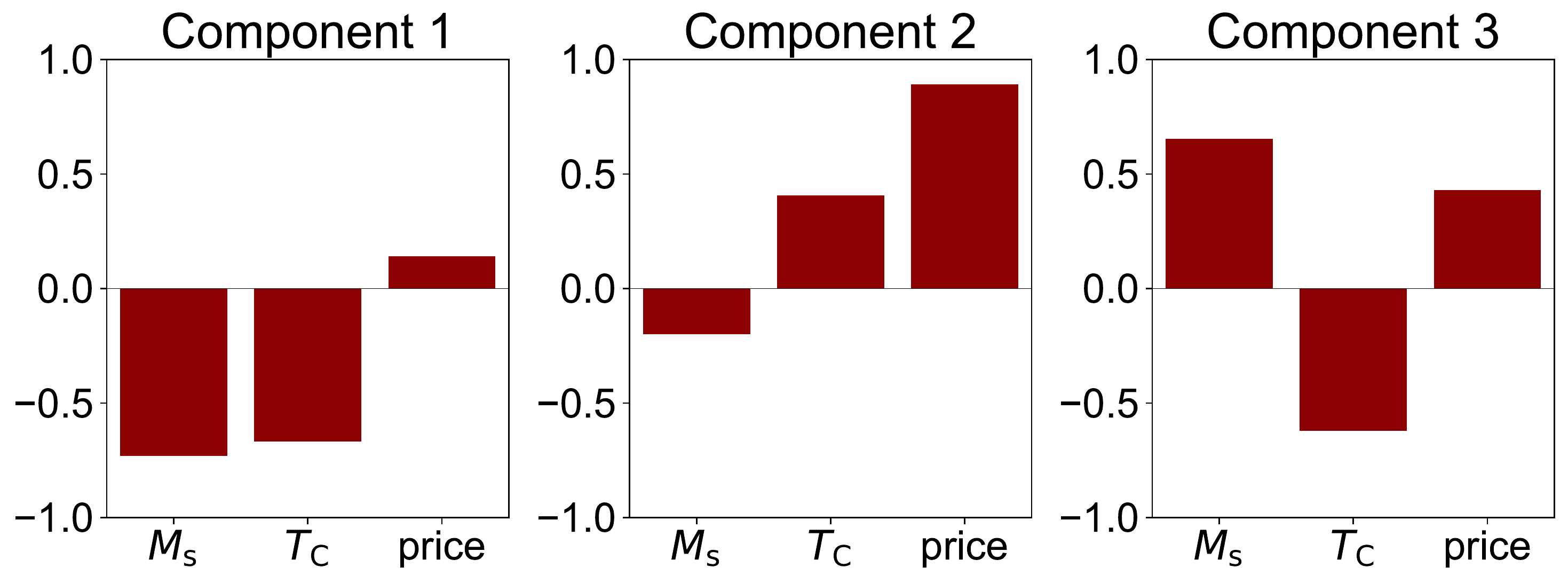}
    \caption{PC1, PC2, and PC3 (left to right) of the target variables
    for (top) the PF and (bottom) all data.
    \label{Fig.targetPCA}}
\end{figure}

PCA can decompose these correlations into 
independent linear combinations of the variables by 
calculating the principal axes of
the estimated covariance.
Figure \ref{Fig.targetPCA}
shows the components of the principal axes,
called principal components (PCs)
with respect to the PF (bottom panels).
We also show the PCs for all data for comparison.

Because we consider three target variables,
there are three principal axes.
We call them PC1, PC2, and PC3 in descending order of
variance along the axes.
The directions of axes are different for the 
PF and all data.
If PCs explain a large part of the variance in the PF
(i.e., if the PFs have large variances along the PC)  
the PCs represent the trade-off among the target variables (Fig.~\ref{Fig.PF}).
Because PC1 for the PF explains 57\% of the variance and 
PC2 for the PF explains 40\%, these two PCs are important.

The trade-offs can be summarized as follows.
PC1 has positive components in the Curie temperature
and price, and an opposite component in magnetization.
Therefore, the Curie temperature increases
at the cost of price and magnetization along the axis.
Similarly, along PC2, the magnetization increases
at the cost of price and Curie temperature.
Both these PCs show a trade-off between the magnetization
and Curie temperature; however, the price index 
has opposite correlations with them. 
This is how the weak correlation seen
in Fig.~\ref{Fig.Parallel} appears in the PCA.

This result is not obtained from the 
the PCA for all data, where PC1, which explains 
90\% of the variance in all data, is almost 
orthogonal to PC1 and PC2 for the PF.
The PF can be used to see
the trade-offs between the target variables clearly.
The PCs for the PF 
resemble those for all data;
PC1 for PF is similar to PC2 for all data,
PC2 for PF is similar to PC3 for all data,
and PC3 for PF is similar to PC1 for all data, although the order of the PCs
are different.
From this observation,
we see that PC1 for all data indicates the direction
that traverses the dominated region, and 
after filtering the non-PF members, the variance 
along the direction is greatly reduced, which is almost
identical to PC3 for the PF.

We now consider controlling the target variables
by changing the input variable
as an example of the control variable analysis in Fig.~\ref{flowchart}.
To analyze the descriptors,
the simple application of PCA
is not meaningful in our case because they are 
controllable variables and determined by an artificial
choice; we choose the candidate systems so that they 
are distributed homogeneously in the descriptor space.
We instead use the PLS2 regression, which searches the linear vector (X-weight)
in the descriptor space that has the largest covariance with 
a vector in the target space (Y-weight) \cite{Holcomb97}.
Let $C$ denote the cross-covariance matrix whose component
is $C_{ij}=\sum_d (x^{(d)}_i - \bar{x}_i)(y^{(d)}_j - \bar{y}_j)$,
where
$x^{(d)}_i$ denotes 
the $d$th data for the $i$th input variable,
$y^{(d)}_j$ denotes 
the $d$th data for the $j$th output variable, and
the bar denotes the mean of the variable.
(Division by a constant is omitted here because it does not change the result).
Then, the singular value decomposition
gives the decomposition,
\begin{equation}
    C=
    \begin{pmatrix}
        {\vec u}_1 &
        \cdots &
        {\vec u}_M
    \end{pmatrix}
    \begin{pmatrix}
        \lambda_1 & &\\
        & \ddots & \\
         & & \lambda_N\\
         & \text{\huge 0}
    \end{pmatrix}
    \begin{pmatrix}
        {}^t {\vec v}_1 \\
        \vdots \\
        \vdots \\
        {}^t {\vec v}_N \\
    \end{pmatrix},
\end{equation}
where $M$ is the input dimension, $N$ is the output dimension,
${\vec u}$ and ${\vec v}$ are sets of
normalized orthonormal bases,
and the diagonal $\lambda$ elements
hold $\lambda_1 \geq \cdots \geq \lambda_N$.
We also assume that $M \geq N$ here for simplicity.
The first X-weight vector is required to be
proportional to ${\vec u}_1$,
and
the first Y-weight vector
is determined by a linear regression of the data
with respect to the values called X-score that are obtained
by projecting the input data to the X-weight axis \cite{Holcomb97}.

In the next step, both the descriptor and target variable space
are reduced by projection to the orthogonal complement of the 
the X- and Y-weight subspace.
For the covariance, $C'$, of reduced data,
the second X- and Y-weights are determined
in the same manner as the first.
This is iterated until one of the spaces is reduced to
zero dimensions.

\begin{figure}
    \centering
    \includegraphics[width=8cm]{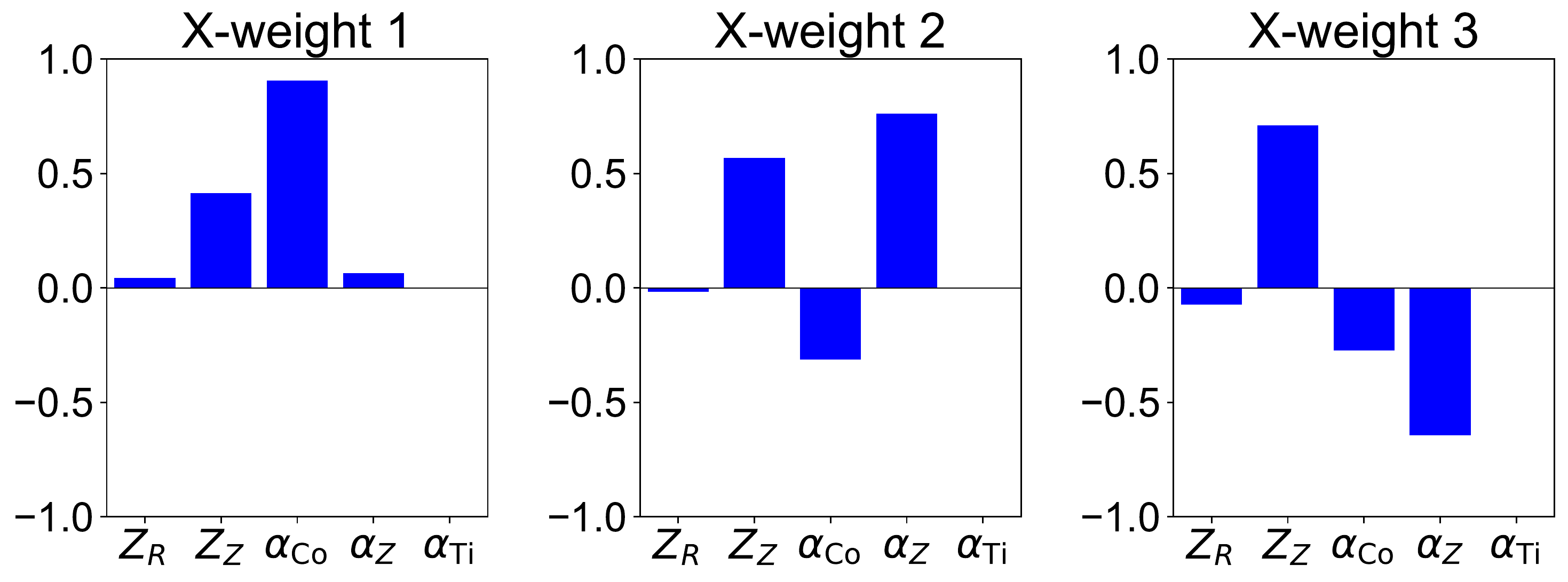}
    \includegraphics[width=8cm]{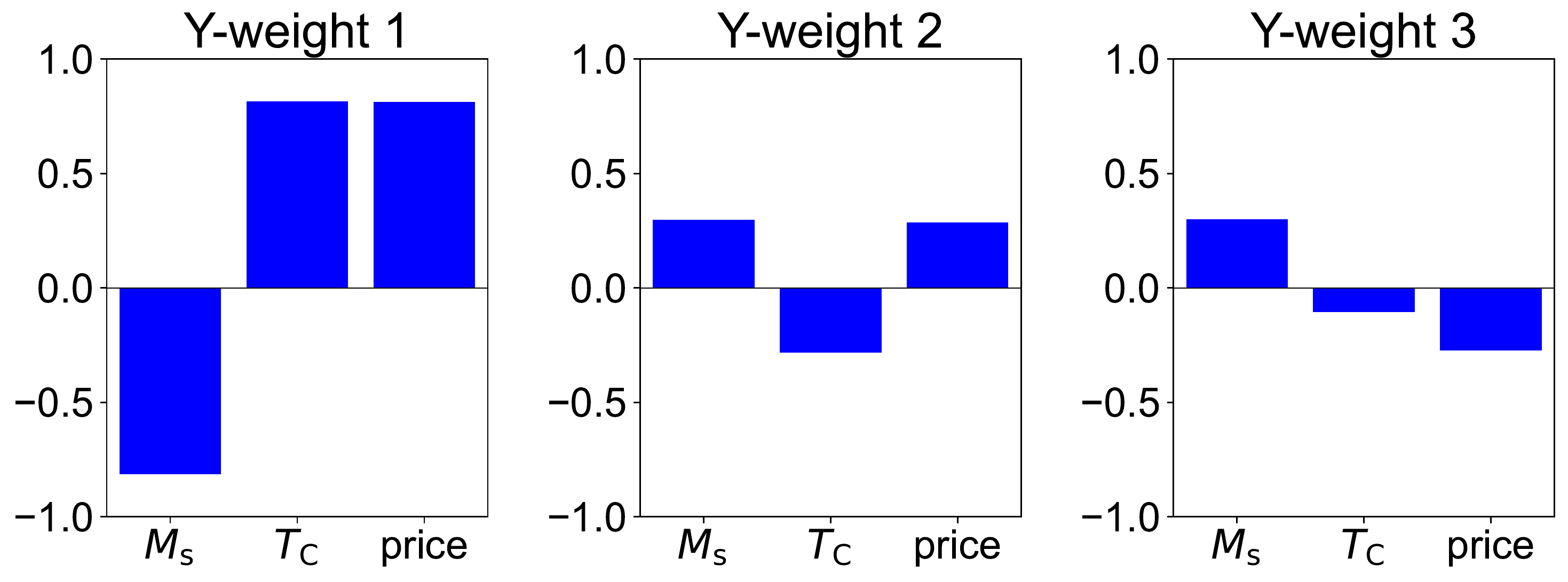}
    \caption{PLS for the PF. Weights for the (top) descriptor and
    (bottom) target variables.\label{Fig.PLSWeights}}
\end{figure}

The Y-weights are not
necessarily identical to the results of the ordinary PCA.
However, a similar set of Y-weights are obtained in our case,
as shown in the bottom panels of Fig.~\ref{Fig.PLSWeights}.
The corresponding X-weights are shown in the top panels,
where we use the atomic numbers of $R$ ($Z_R$) and $Z$ ($Z_Z$),
and the concentrations of Co ($\alpha_\mathrm{Co}$), 
$Z$ ($\alpha_\mathrm{Z}$), and
Ti ($\alpha_\mathrm{Ti}$)
as descriptor (explanatory) variables.
This choice of descriptors is based on our previous study,
in which we referred to this type of descriptor as \#9 \cite{Fukazawa19c}.

The first Y-weight
is similar to PC1 for PF in Fig.~\ref{Fig.targetPCA},
and the corresponding X-weight
has a large weight for Co (Fig.~\ref{Fig.PLSWeights}). The analysis rediscovers 
the efficiency of Co in enhancing the Curie temperature,
which is already well-known, with the sacrifice of 
magnetization and cost.
Because none of the PF systems contain Ti, the weight for Ti is zero
for all components in Fig.~\ref{Fig.PLSWeights}.

The second Y-weight is 
similar to PC1 for the PF in Fig.~\ref{Fig.targetPCA},
which represents a direction
toward the increase in magnetization at the cost of the price
and Curie temperature.
The corresponding
X-weight describes the choice of Dy as
the $Z$ element, the increase of the $Z$ element, and 
the reduction of Co. 
Although the enhancement of magnetization  by the reduction of Co is foreseeable,
the contribution of Dy to magnetization is unexpected.

\begin{figure}
    \centering
    \includegraphics[width=8cm]{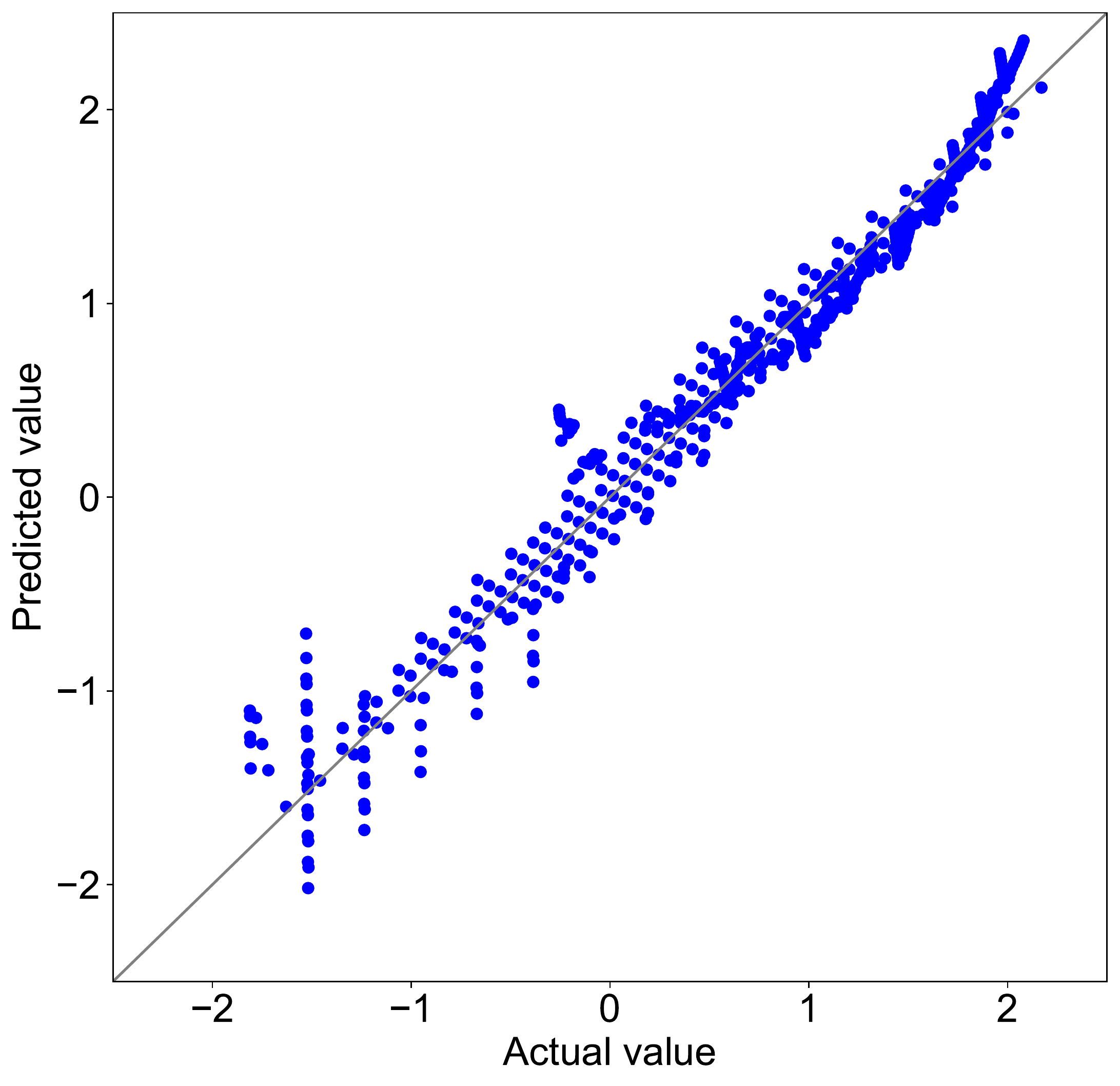}
    \caption{Actual versus predicted values with the PLS2 regression.\label{Fig.PLSfit}}
\end{figure}

To see the validity of the PLS regression,
we predict the magnetization,
Curie temperature, and price index with
the data projected to the X and Y-weight axes,
and
compare the standardized values of them with the actual values
in Fig.~\ref{Fig.PLSfit}.
The coefficient of determination is 0.947, which 
shows the validity of the linear model for 
describing the distribution in the PF.

\subsection{MBO}
\label{SS:MBO}
\begin{figure}
    \centering
    \includegraphics[width=8cm]{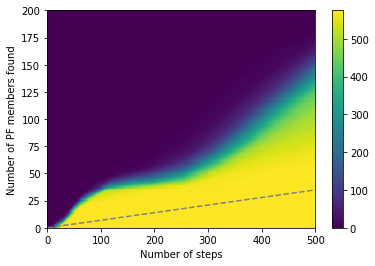}
    \caption{Step number versus number of true PF members found with MBO.\label{Fig.foundPF}
    The color map shows the frequency of obtaining a score higher than the 
    vertical value in 575 sessions.
    The dashed line shows the analytic results with random sampling.}
\end{figure}

Although the analysis using the PF is powerful,
obtaining the PF can be a difficult;
in the present case, the PF consists of 253 members
from the 3630 data points.
First, we see how 
MBO
is efficient in finding these true PF members.
Because the MBO process uses a stochastic process, 
we need a statistical analysis in the performance evaluation.
Figure~\ref{Fig.foundPF} shows the number of 
true PF members found in the MBO search
to determine how efficient MBO is.

The color map in the figure shows the frequency of sessions
that obtained a score higher than the value on the vertical axis
within the number of steps indicated on the horizontal axis.
The dashed line shows the mean score for random sampling.
Although the MBO search is much more efficient than random sampling, 
it cannot find all the 253 PF members within 500 steps.

However,
a tentative PF from
the MBO session can serve as an 
approximated PF, from which the features of the true PF
can be understood.
To examine the validity of the approximated PFs,
we construct a model for the target variables
from tentative PFs with PLS2, 
and see how the model describes the true PF
in terms of coefficient of determination.

\begin{figure}
    \centering
    \includegraphics[width=8cm]{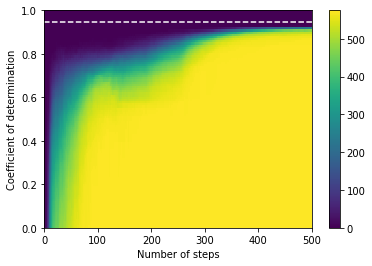}
    \caption{Step number versus the coefficient
    of determination (COD) with PLS2\label{Fig.CODPLS}.
    The color map shows the frequency of obtaining
    a COD value higher than the 
    vertical value in 575 sessions.
    The dashed line shows the COD value (0.947)
    with the true PF.}
\end{figure}

Figure~\ref{Fig.CODPLS} shows the frequency of the 
sessions that obtained a coefficient of determination value larger
than the value on the 
vertical axis within the steps indicated on
the horizontal axis.
In this case, 400 steps seems sufficient to 
generate a model that describes the 
features of the true PF
when the number of the true PF members is 253.
Therefore, this MBO scheme seems efficient in 
obtaining an approximate PF that can represent the true
PF behaviors.

\section{Conclusion}
We proposed a scheme for analyzing the multi-objective 
correlation and the trade-offs among the target variables (Fig.~\ref{flowchart}).
To determine the usefulness of the analyzer part of the framework, 
we performed PCA and PLS analysis
for magnet compounds 
\ch{(R_{$1-\alpha$}$Z$_{$\alpha$})
(Fe_{$1-\beta$}Co_{$\beta$})_{$12-\gamma$}Ti_{$\gamma$}}
to extract information for searching for
PF materials from first-principles data.
The variance along the PF
was characterized by PC1, which was
controlled mainly by the introduction of Co,
and by PC2,
which was controlled mainly by
simultaneous introduction of Dy and reduction of Co.

To demonstrate the efficiency of the scheme for obtaining 
an approximate PF, which is described as a loop 
before the analysis in Fig.~\ref{flowchart},
we conducted a performance evaluation with 
MBO.
We showed that MBO obtained PF members much
more efficiently than random sampling,
and generated 
an approximate PF that 
adequately represented the true PF.
We showed that
a model could be constructed
that described the true PF well
from an approximate PF obtained
with 400 first-principles calculations, 
where all the candidates consisted of 3630 systems.

\section*{Acknowledgment}
This work was supported by 
%
%
%
%
%
the ``Program for Promoting Researches on the Supercomputer Fugaku''
(DPMSD, Project ID: JPMXP1020200307) by MEXT.
The calculations were conducted in part using the facilities of the Supercomputer Center at
the Institute for Solid State Physics, University of Tokyo.
%
%
%
%
%
%



\bibliography{sn}
\end{document}